\documentclass[12pt, a4paper]{article}
\usepackage{hyperref}
\usepackage{geometry}
\usepackage{graphicx}
\usepackage{subcaption}
\usepackage[numbers]{natbib}
\usepackage{amssymb}                                                                                                                                                                                                                                                                                                                                                                                                                                                                                                                                                                                                                                                                                                                                                                                                                                                                                                                                                                                                                                                                                                                                                                                                                                                                                                                                                                                                                                                                                                                                                                                                                                                                                                                                                                                                                                                                                                                                                                                                                                                                                                                                                                                                                                                                                                                                                                                                                                                                                                                                                                                                                                                                                                                                                                                                                                                                                                                                                                                                                                                                                                                                                                                                                                                                                                                                                                                                                                                                                                                                                                                                                                                                                                                                                                                                                                                                                                                                                                                                                                                                                                                                                                                                                                          
\usepackage{amsmath}
\usepackage{slashed}
\usepackage{anyfontsize}
\usepackage{bm}
\usepackage{slashed}
\usepackage{multirow}
\usepackage{float}
\usepackage[font=small]{caption}
\usepackage{listings}
\usepackage{simpler-wick}
\usepackage{physics}
\usepackage{color}
\definecolor{codegreen}{rgb}{0,0.5,0}
\definecolor{codeblue}{rgb}{0,0,0.5}
\definecolor{maroon}{rgb}{0.675,0.1,0.1}
\usepackage{upgreek}
\usepackage{tensor}

\usepackage[utf8]{inputenc}
\usepackage[english]{babel}
\usepackage{amsthm}
\usepackage{slashed}
\usepackage{xcolor}
\definecolor{ForestGreen}{RGB}{0, 176, 80}
\usepackage{hyperref}
\newgeometry{lmargin=0.9in, rmargin=0.9in, tmargin=1in, bmargin=1in}
\title{On the Particle and Field nature of $\gamma^\mu$ matrices in the Dirac Equation and the Nature's intrinsic fifth force}
\author{B.T.T.Wong\footnote{The University of Hong Kong,  u3500478@connect.hku.hk}}
\date{}

\theoremstyle{definition}

\begin{document}
\maketitle
\begin{abstract}
The Dirac equation is a cornerstone of modern particle physics, which integrates special relativity and quantum mechanics into a consistent framework, yielding the prediction of electron and its antiparticle counterpart, positron. The Dirac equation also lays the foundation of quantum electrodynamics, such that QED phenomenon is supported by fundamental Dirac Algebras calculation. In this article, we will introduce new perspectives of the $\gamma^\mu$ matrix in the Dirac Algebra, by realizing the $\gamma^\mu$ matrices are actual formal quantum fields, the excitation of $\gamma^\mu$ fields correspond to a new particle with both boson and fermion nature. Thus, we show that $\gamma^{\mu}$ is a particle in nature, and can be referred as the nature's intrinsic fifth force. The  $\gamma^\mu$ field also serves as the boson-fermion connector in QED interaction.
\end{abstract}
\section{Introduction}
The Standard Model (SM) of particles are mainly classified into spin-0 and spin -1 Gauge Bosons, which are described by the excitation of gauge fields; and spin 1/2 fermions are described by quantization of Dirac spinor fields. The fermions include leptons and quarks, which are the matter building blocks; and the spin-one gauge bosons include photon, which is responsible for the electromagnetic interaction; $W^{\pm}$ and $Z$ bosons, which are responsible for the weak interaction; and eight gluons which are force carriers of the strong interaction. Finally, the spin-zero Higgs boson, the only scalar particle in SM, is responsible for accounting masses of the three electroweak bosons, as well as the fermion masses through the process of spontaneous symmetry breaking \citep{Higgs1, Higgs2, Englert, Weinberg, Salam, Kibble}. Gravity, on the other hand, is manifested as spacetime geometry which is governed by Einstein theory of general relativity. All together, the electromagnetic force, strong force, weak force and gravity constitute the four fundamental forces in nature. 

The Dirac equation, proposed by Dirac in 1928 \citep{Dirac}, describes the dynamics of a relativistic fermion compatible with special relativity and quantum mechanics, reads as follows
\begin{equation}
(i\gamma^\mu \partial_{\mu} -m) \psi = 0\,,
\end{equation} 
where the $\gamma^\mu$s are $4\times 4$ matrices object that satisfies the Dirac algebra,
\begin{equation}
\{\gamma^\mu , \gamma^\nu \} = 2\eta^{\mu\nu}\pmb{1}_{4\times 4}
\end{equation}
with $\eta^{\mu\nu}$ the inverse of flat Minkowski metric. Mathematically, the set of gamma matrices $\gamma^\mu =\{\gamma^0, \gamma^1, \gamma^2 ,\gamma^3  \}$ generates basis representation of the Clifford Algebra $\mathrm{Cl_{1,3}}(\mathbb{R})$, and its commutator $[\gamma^\mu , \gamma^\nu]=\gamma^\mu\gamma^\nu - \gamma^\nu\gamma^\mu$ is the basis for the Lorentz algebra which facilitate three spatial rotations and three Lorentz boosts \cite{Peskin}. Physically, we know that $ie\gamma^\mu$ acts as a Feynman vertex in the momentum space. However, generally speaking, the intrinsic physical meaning of $\gamma^\mu$ matrix is not very clearly known.

In this paper, we will novelly realize constant gamma matrices $\gamma^\mu$ as quantum field particle excitations in physics means. Secondly, we will explicitly demonstrate the Wick's theorem of $\gamma^\mu $matrix. Then finally, we will work out the formalism of gamma fields in differential geometry means, and we find that the electromagnetic field tensor $F^{\mu\nu}$ can be represented by gamma matrices. Then the integral formalism of gamma matrices is also studied.

\section{Constant classical background field $\gamma^\mu$}
In this section, we will give an intensive study of the well-known subject $\gamma^\mu$ matrix in the Dirac equation. Explicitly, we would like to investigate the physical meaning of the gamma matrix, beyond the known fact of just the basis of Clifford algebra. 
 
First of all, the $\gamma^\mu$ matrix transforms like a vector field, and it is a constant vector field,
\begin{equation}
\gamma^{\prime\mu} = \Lambda^{\mu}_{\nu} \gamma^\nu \,, 
\end{equation}
where $\Lambda^{\mu}_{\nu}$ is the Lorentz transformation. 

Since $\gamma^\mu$ is a constant vector field, it admits a global transformation instead of a local transformation. We know that it globally transforms as
\begin{equation}
V \equiv \Lambda_{\frac{1}{2}} = e^{-\frac{i}{2}\omega_{\mu\nu}S^{\mu\nu}}\,,
\end{equation}
where $\omega_{\mu\nu}$ is the infinitesimal global parameter and $S^{\mu\nu} =\frac{i}{4}[\gamma^\mu, \gamma^\nu]$ is the generator of the Lorentz algebra for the Lorentz group $SO(1,3)$. And we have the following set of global transformation rule under global SO(1,3) Lorentz group
\begin{equation} \label{eq:T1}
\begin{cases}
\gamma^{\prime\mu} = V^{-1}\gamma^\mu V\\
\psi^\prime = V\psi \,.
\end{cases}
\end{equation}
Let's take a reference to local transformation of non-abelian gauge field \cite{Yang}
\begin{equation} \label{eq:t2}
\begin{cases}
A^{\prime \mu} = UA^{\mu}U^{-1} +\frac{i}{g}U\partial^\mu U^{-1}\\
\psi^\prime = U\psi
\end{cases}
\end{equation}
where $U(x) = e^{i\alpha^a(x)t^a}$ is an element of a SU(N) Lie group. There is a remarkable similarity between \ref{eq:T1} and \ref{eq:t2}. The only difference is the former is global and the latter is local, and the former one belongs to global non-abelian SO(1,3) group while the latter belongs to a general non-abelian Lie group. Notice that the second term of the first equation in \label{eq:t1} vanishes because we have a global transformation, explicitly we should write
\begin{equation}
\gamma^{\prime\mu} = V^{-1}\gamma^\mu V + \frac{i}{g^\prime} V^{-1}\partial^\mu V
\end{equation} 
but $\partial^\mu V =0$ as $\partial^{\mu}\omega_{\alpha\beta} =0$ given that $\omega_{\mu\nu}$ is independent of spacetime.

The only difference is that $V$ is non-unitary due to the fact that the SO(1,3) group generators are non-hermitian, while $U$ unitary as the Lie group generators are unitary. 

Therefore, we can see that the $\gamma^\mu$ matrix behaves like a non-abelian gauge field, at least classically. We we investigate the tensor notation, one finds $\gamma^{\mu}_{ab}$ versus $A^{\mu}_{ij}\equiv A^{\mu a}t^{a}_{ij}$. Similarly the field strength is defined by
\begin{equation}
F_{\mu\nu} = \partial_{\mu}\gamma_{\nu} - \partial_{\nu}\gamma_\mu + \gamma_{\mu}\gamma_{\nu} -\gamma_{\nu}\gamma_{\mu} = -4iS_{\mu\nu}\,,
\end{equation}
contrasting to non-abelian gauge field strength
\begin{equation}
F_{\mu\nu} = \partial_{\mu}A_{\nu} - \partial_{\nu}A_\mu + A_{\mu}A_{\nu} -A_{\nu}A_{\mu} \,.
\end{equation}
Compactly,
\begin{equation}
F = d\gamma + \gamma\wedge\gamma =\gamma\wedge\gamma
\end{equation}
and
\begin{equation}
F = dA + A \wedge A \,.
\end{equation}
The Lagrangian is given by
\begin{equation}
\mathcal{L} = -16\mathrm{Tr} S_{\mu\nu}S^{\mu\nu} = -16\times -192 = 3072\,
\end{equation}
which is a constant. The number $-192$
denotes the spin field strength. 

In non-Abelian field strength, it transforms as
\begin{equation}
F_{\mu\nu}^\prime = UF_{\mu\nu}U^{-1}\,.
\end{equation}
We can show that the spin tensor also transforms globally in this way:
\begin{equation}
\begin{aligned}
S_{\mu\nu}^\prime &= 
\frac{i}{4}[\gamma_{\mu}^\prime,\gamma_{\nu}^\prime ] \\
&=\frac{i}{4}[\gamma_\mu^\prime \gamma_\nu^\prime -\gamma_\nu^\prime \gamma_\mu^\prime] \\
&=\frac{i}{4}[\Lambda_{\frac{1}{2}}^{-1}\gamma_{\mu}\Lambda_{\frac{1}{2}} \Lambda_{\frac{1}{2}}^{-1}\gamma_{\nu} \Lambda_{\frac{1}{2}} -\Lambda_{\frac{1}{2}}^{-1}\gamma_{\nu}\Lambda_{\frac{1}{2}} \Lambda_{\frac{1}{2}}^{-1}\gamma_{\mu} \Lambda_{\frac{1}{2}} ] \\
&= \Lambda_{\frac{1}{2}}^{-1} \bigg( \frac{i}{4}[\gamma_{\mu}\gamma_{\nu}- \gamma_{\nu}\gamma_{\mu}]\bigg) \Lambda_{\frac{1}{2}}\\
&=\Lambda_{\frac{1}{2}}^{-1} S_{\mu\nu} \Lambda_{\frac{1}{2}} \,. 
\end{aligned}
\end{equation}
Therefore, the gamma field behaves as global gauge field.

\section{Canonical Quantization} 
 It is known that fermionic spinor can be quantized by equal-time anticommutation relation,
 \begin{equation} \label{eq:EATCR}
 \{ \psi_a(\pmb{x}), \psi_b^\dagger(\pmb{y})\} = \delta^3 (\pmb{x} - \pmb{y})\delta_{ab} \,.
 \end{equation}
and
\begin{equation}
\{ \psi_a(\pmb{x}), \psi_b(\pmb{y})\} =  \{ \psi^\dagger_a(\pmb{x}), \psi^\dagger_b(\pmb{y})\} = 0
\end{equation}
And indeed we find that the constant gamma field is naturally quantized, by the Dirac algebra of the $\gamma^\mu$ field and its complex conjugate $\gamma^{\mu\dagger}$,
\begin{equation}
\{\gamma^{\mu}_{ac}, \gamma^{\nu\dagger}_{cb}\} =2\delta^{\mu\nu}\delta_{ab}\,
\end{equation}
The factor of 2 is just from normalization. If we consider in 3 spatial dimensions, we obtain an expression in analogy to the anti-commutation relation of the fermonic field in \ref{eq:EATCR},
\begin{equation}
\{\gamma^{i}_{ab}, \gamma^{j\dagger}_{bc}\} =2\delta^{ij}\delta_{ac}\,,
\end{equation}  
and also
\begin{equation}
\{\gamma^{i}_{ab}, \gamma^{j}_{bc}\} =\{\gamma^{i\dagger}_{ab}, \gamma^{j\dagger}_{bc}\} =0
\end{equation}
for $i\neq j$.

Therefore, we see that $\mu$ has the role of $x$ and $\nu$ has the role of $y$, and in 3 dimension $i$ has the role of $\pmb{x}$ and $j$ has the role of $\pmb{y}$. 

Next we would like to carry out canonical quantization of the $\gamma^\mu$ field using fermonic creation and annihilation operators. Let's take a look on the canonical quantization of gauge field first. The quantized U(1) gauge field is
\begin{equation}
\hat{A}^{\mu}(\pmb{x}) = \int \frac{d^3 \pmb{p}}{\sqrt{(2\pi)^3}\sqrt{2E_{\pmb{p}}}} \sum_{\lambda}(\epsilon^\mu_\lambda \hat{a}_{\pmb{p}}  + \epsilon^{*\mu}_\lambda \hat{a}_{-\pmb{p}}^\dagger ) e^{i\pmb{p}\cdot \pmb{x}} \,, 
\end{equation}
where $\epsilon^\mu_\lambda$ is the polarization vector. For fermions, we need two different creation and annihilation operators. We ansatz the quantization of $\gamma^\mu$ field as follow because the field is independent of position, and the momenta are discrete modes:
\begin{equation}
\hat{\gamma}^{\mu}_{ab} = \sum_{\pmb{p}} \sum_\lambda \frac{1}{\sqrt{(2\pi)^3}} (\epsilon^\mu_\lambda \theta_{ab} \hat{b}_{\pmb{p}} +\epsilon^{*\mu}_{\lambda} \theta^*_{ab} \hat{c}^\dagger_{-\pmb{p}}) \,,
\end{equation}
and
\begin{equation}
\hat{\gamma}^{\mu\dagger}_{bc} =  \sum_{\pmb{p}} \sum_\lambda \frac{1}{\sqrt{(2\pi)^3}} (\epsilon^{*\mu}_\lambda \theta_{bc} \hat{b}_{-\pmb{p}}^\dagger +\epsilon^{\mu}_{\lambda} \theta_{ab} \hat{c}_{\pmb{p}}) \,,
\end{equation}
 We ansatz the constant tensor to satisfy the following normalization
\begin{equation} 
\theta_{ab} \theta_{bc}^* = \delta_{ac} \quad\text{and}\quad \theta_{ab}^* \theta_{bc} = \delta_{ac}  
\end{equation}
And the creation and annihilation operators satisfies the following anti-commutation relation
\begin{equation}\label{eq:acom}
\{ \hat{b}_{\pmb{p}},  \hat{b}_{\pmb{p}^\prime}^\dagger\} = (2\pi)^3 \delta_{\pmb{p} \pmb{p}^\prime} \quad \text{and}\quad \{ \hat{c}_{\pmb{p}},  \hat{c}_{\pmb{p}^\prime}^\dagger\} = (2\pi)^3 \delta_{\pmb{p} \pmb{p}^\prime}\,,
\end{equation}
while all other anti-commutation relations are zero. Now, we compute the anti-commutator of the field:
\begin{equation}
\begin{aligned}
\{\hat{\gamma}^\mu_{ac}, \hat{\gamma}^{\mu^\dagger}_{cb}\} &=\sum_{\pmb{p}}\sum_{\pmb{p}^\prime}\sum_{\lambda}\sum_{\rho}\frac{1}{(2\pi)^3}(\epsilon^\mu_\lambda\epsilon^{*\nu}_\rho \theta_{ac}\theta^*_{cb}\{\hat{b}_{\pmb{p}} , \hat{b}_{-\pmb{p}^\prime}^\dagger \}+   \epsilon^{*\mu}_\lambda\epsilon^\nu_\rho \theta_{ac}^*\theta_{cb}\{\hat{c}^\dagger_{\pmb{-p}} , \hat{c}_{\pmb{p}^\prime} \}  )\\
&=\sum_{\pmb{p}}\sum_{\pmb{p}^\prime}\sum_{\lambda}\sum_{\rho}\frac{1}{(2\pi)^3}(\epsilon^\mu_\lambda\epsilon^{*\nu}_\rho \theta_{ac}\theta^*_{cb}\{\hat{b}_{\pmb{p}} , \hat{b}_{-\pmb{p}^\prime}^\dagger \}+   \epsilon^{*\mu}_\lambda\epsilon^\nu_\rho \theta_{ac}^*\theta_{cb}\{\hat{c}_{\pmb{p}^\prime} , {\hat{c}}^\dagger_{\pmb{-p}}  \}  )\\
&=\sum_{\pmb{p}}\sum_{\pmb{p}^\prime}\sum_{\lambda}\sum_{\rho}\frac{1}{(2\pi)^3}(\epsilon^\mu_\lambda\epsilon^{*\nu}_\rho \delta_{ab}(2\pi)^3 \delta_{\pmb{p}, -\pmb{p}^\prime}+  (2\pi)^3 \epsilon^{*\mu}_\lambda\epsilon^\nu_\rho \delta_{ab}\delta_{\pmb{p}^\prime, -\pmb{p}} )\\
&=\sum_{\lambda}\sum_{\rho}(\epsilon^\mu_\lambda\epsilon^{*\nu}_\rho+\epsilon^{*\mu}_\lambda\epsilon^{\nu}_\rho)\delta_{ab}\\
&=\sum_\lambda  (\epsilon^\mu_\lambda\epsilon^{*\nu}_\lambda +\epsilon^{*\mu}_\lambda\epsilon^{\nu}_\lambda)\delta_{ab} \\
&= -2\eta^{\mu\nu}\delta_{ab}
\end{aligned}
\end{equation}
where in the third line we have used the fact that
\begin{equation}
\sum_{\pmb{p}\pmb{p}^\prime}\delta_{\pmb{p}, -\pmb{p}^\prime} =1
\end{equation}
as only the term when $\pmb{p} = \pmb{p}^\prime = 0$ survives. In the fifth line, we have used the fact from polarization sum of massless particle that
\begin{equation}
\sum_{\lambda} \epsilon^\mu_\lambda\epsilon^{*\nu}_\lambda =\sum_{\lambda}\epsilon^{*\mu}_\lambda\epsilon^{\nu}_\lambda = -\eta^{\mu\nu}
\end{equation}
As the canonical quantization only concerns about the spatial dimensions, this suffices to give
\begin{equation}
\boxed{
\{\hat{\gamma}^i_{ac}, \hat{\gamma}^{j\dagger}_{cb}\} =2\delta^{ij}\delta_{ab}
}
\end{equation}
and it is easy to show that
\begin{equation}
\boxed{
\{\hat{\gamma}^i_{ac}, \hat{\gamma}^{j}_{cb}\} =\{\hat{\gamma}^{i\dagger}_{ac}, \hat{\gamma}^{j\dagger}_{cb}\}=0 } \,. 
\end{equation}
as all other anti-commutation relations other than \ref{eq:acom} are zero.

Therefore we see that the constant field $\gamma^{\mu}$ behaves like both a boson and a fermion, for which it satisfies the global gauge transformation rule and its quantization satisfies anti-commutation relationship. This special property can be in fact demonstrated through the QED interaction
\begin{equation}
\mathcal{L}_{\mathrm{QED, interaction}} = -e\bar{\psi}_a \gamma^{\mu}_{ab}A_{\mu} \psi_b \,,
\end{equation}
so the gamma matrix field $\gamma^{\mu}$ connects both the fermion (Dirac spinor) and the gauge boson $A_{\mu}$. This makes a lot of sense that the gamma field contains both bosonic and fermonic properties as it connects to both types of particle. We term the gamma field as the boson-fermion connector. Furthermore, the gamma field $ie\gamma^\mu$ is known as the QED vertex, in which it connects two spinor fermions and one boson. 

\section{Correlation function of Gamma $\gamma^{\mu}$ fields}
As in the last section we have seen that the $\gamma^\mu$ matrix serve as constant quantum fields, therefore $\gamma^\mu$ itself is considered as a particle with undefined spin, as it is neither a fermion nor a boson. Now we will work out the $n-$point Green's function of the $\gamma^\mu$ fields. 

The $n-$point Green's function of the $\gamma^\mu$ fields is simply the trace of $n$ gamma matrices. The reason will be apparent when we work out some examples. First notice that the trace of odd number of gamma matrices is zero, and the $n-$point correlation function of fields for odd $n$ is zero. 

For example, we have
\begin{equation}
\langle 0|\gamma^\mu |0\rangle = \mathrm{Tr}\gamma^\mu = 0 \,.
\end{equation}
The vacuum expectation value of fields is zero
\begin{equation}
\langle 0| \hat{\psi} (x) |0\rangle = 0
\end{equation}
Next
\begin{equation}
\langle 0|\hat{\gamma}^\mu \hat{\gamma}^\nu|0\rangle = \mathrm{Tr}(\gamma^\mu\gamma^\nu )= 4\eta^{\mu\nu} \,.
\end{equation}
And we see that
\begin{equation}
\langle 0|\mathrm{T}\hat{\phi}(x)\hat{\phi}(y)|0\rangle = \Delta_F (x-y)
\end{equation}
So we can see that $\mu$ maps to $x$ and $\nu$ maps to $y$.
Furthermore see that
\begin{equation} \label{eq:tr}
\langle 0 |\mathrm{T}\hat{\gamma}^{\mu}\hat{\gamma}^{\nu}\hat{\gamma}^{\rho}\hat{\gamma}^{\sigma} |0\rangle =\mathrm{Tr}(\hat{\gamma}^{\mu}\hat{\gamma}^{\nu}\hat{\gamma}^{\rho}\hat{\gamma}^{\sigma}) = 4(\eta^{\mu\nu}\eta^{\rho\sigma} -\eta^{\mu\rho}\eta^{\nu\sigma} +\eta^{\mu\sigma}\eta^{\nu\rho}) \,,
\end{equation}
so this is just the contraction over the indices. The factor of 4 is unimportant and can be normalized later. When compare to the contraction of normal fields
\begin{equation}
\begin{aligned}
&\quad\langle 0|\hat{\phi}(x_1)\hat{\phi}(x_2)\hat{\phi}(x_3)\hat{\phi}(x_4)|0\rangle\\
& = \Delta_F(x_1 - x_2)\Delta_F(x_3 - x_4) + \Delta_F(x_1 - x_3)\Delta_F(x_2 - x_4) + \Delta_F(x_1 - x_4)\Delta_F(x_2 - x_3) \,.
\end{aligned}
\end{equation}
So we map $\mu$ to $x_1$, $\nu$ to $x_2$, $\rho$ to $x_3$ and  $\sigma$ to $x_4$. 
Everything is similar except for the minus sign of the second term and the front factor of 4.

Let's investigate \ref{eq:tr} in detail diagramatically
\begin{figure}[H]
\centering 
\includegraphics[scale=0.6]{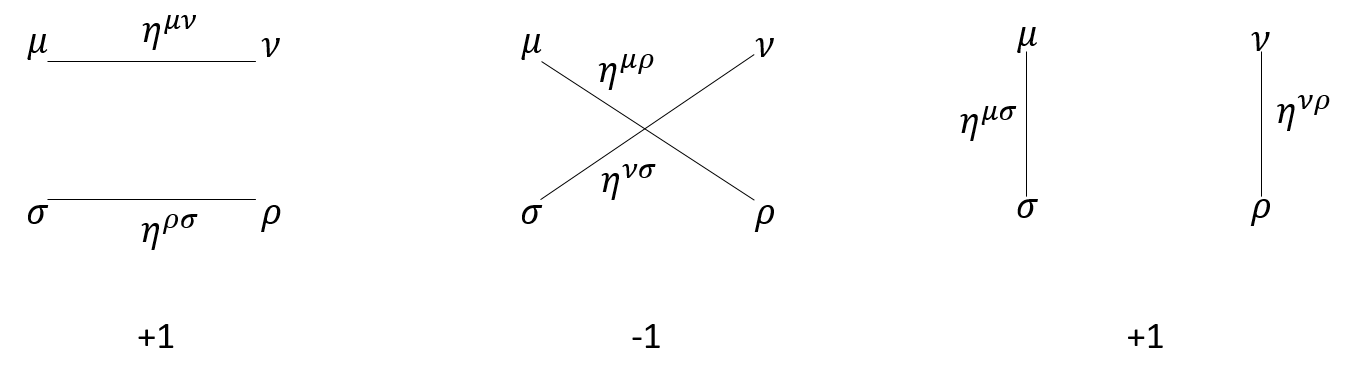}
\caption{}
\label{trace}
\end{figure}
The sign difference can be related to the topology of these graphs. Suppose we join the initial and end points, which is as follows:
\begin{figure}[H]
\centering 
\includegraphics[scale=0.6]{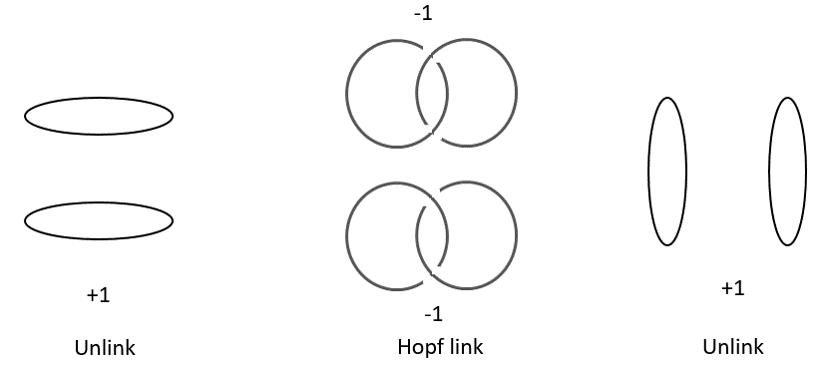}
\caption{}
\label{topo6}
\end{figure}
The cross diagram is degenerated into two diagrams with different handiness. Therefore, we can see that terms with same sign have the same topology. The distinction of the cross diagram and the uncross diagram can be seen alternatively by the contraction index. Define the index indication function $f(\mu_i , \mu_j) = i +j$ where $i,j$ = $1,2,3,4$. Thus for each contraction we have the contraction index pair $(f(\mu_i , \mu_j),f(\mu_k , \mu_l)) = (i+j , k+l)$, and $f(\mu_i , \mu_j) + f(\mu_k , \mu_l)=(i+j)+(k+l) = 10$. For example, the index pair of $\eta^{\mu\nu}\eta^{\rho\sigma}$ is $(1+2 , 3+4) = (3,7)$. The index pair of $\eta^{\mu\sigma}\eta^{\nu\rho}$ is $(1+4 , 2+3) = (5,5)$. These two contractions have odd index pairs, therefore they are in the same topology. We see that the index pair of $\eta^{\mu\rho}\eta^{\nu\sigma}$ is $(1+3,2+4) = (4,6)$, which is even. This distinguishes the cross diagram from the other two. 

\subsection{$N$-point correlation function of $\gamma^\mu$ fields and Wick's theorem}
Now we study the rigorous definition of $N$-point correlation function of gamma constant quantum fields. From the above intuitive demonstration, we have already shown that the $N$-point correlation function is given by the the sum over all possible contractions. However, we still leave with the sign problem, in particularly the $-\eta^{\mu\rho}\eta^{\nu\sigma}$ term. Now, we will rigorously deal with all these issues. First let's use again the 4-point example. We reconsider the contraction as follows.
Since $\mu\neq\nu$, we have $\gamma^{\mu}\gamma^\nu = -\gamma^\nu\gamma^\mu$. 
Consider the canonical way for the contraction. We want to fix it as the contraction of the first two pairs. For notation convenience, we will drop the hat symbol.
\begin{equation}
\begin{aligned}
\langle 0 |\mathrm{T}\gamma^{\mu}\gamma^{\nu}\gamma^{\rho}\gamma^{\sigma} |0\rangle &= 
\wick{
\langle 0|\mathrm{T} \c1{\gamma}^{\mu} \c1{\gamma}^{\nu}\c1{\gamma}^{\rho} \c1{\gamma}^{\sigma} |0 \rangle +(-1)\langle 0|\mathrm{T} \c1{\gamma}^{\mu} \c1{\gamma}^{\rho}\c1{\gamma}^{\nu} \c1{\gamma}^{\sigma} |0 \rangle + (-1)^2 \langle 0|\mathrm{T} \c1{\gamma}^{\mu} \c1{\gamma}^{\sigma}\c1{\gamma}^{\nu} \c1{\gamma}^{\rho} |0 \rangle
  } \\
&=\frac{1}{16}\mathrm{Tr}(\gamma^\mu\gamma^\nu)\mathrm{Tr}(\gamma^\rho \gamma^\sigma) -\frac{1}{16}\mathrm{Tr}(\gamma^\mu\gamma^\rho)\mathrm{Tr}(\gamma^\nu \gamma^\sigma) + \frac{1}{16}\mathrm{Tr}(\gamma^\mu\gamma^\sigma)\mathrm{Tr}(\gamma^\nu\gamma^\rho) \\
&=   \eta^{\mu\nu}\eta^{\rho\sigma} -\eta^{\mu\rho}\eta^{\nu\sigma} +\eta^{\mu\sigma}\eta^{\nu\rho}
\end{aligned}
\end{equation}
where for each contraction we normalized by the dimension $D=4$.

In general, we can find that the $N$-point correlation function of the gamma field can be formalized by our generalized Wick's theorem as follow. 
\begin{equation}
\langle 0| \mathrm{T}\gamma^{\mu_1}\gamma^{\mu_2}\cdots \gamma^{\mu_N}|0\rangle =\frac{1}{D^{N/2}} \sum_{\substack{i_1 ,i_2 ,\cdots, i_N \in S \\i_1 < i_2 <\cdots < i_N }} \Theta(\mu_{i_1},\mu_{i_2},\cdots ,\mu_{i_N} )\langle 0|\gamma^{\mu_{i_1}}\gamma^{\mu_{i_2}}\cdots \gamma^{\mu_{i_N}}|0\rangle \,.
\end{equation}
Here $S$ denotes the set of all possible moves under the time order constraint $i_1 < i_2 <\cdots < i_N$.
$\Theta(\mu_{i_1},\mu_{i_2},\cdots ,\mu_{i_N} )$ is a function of indices which gives factors of $(-1)^m$ where $m$ is the number of moves. Explicitly
\begin{equation}
\Theta(\mu_{i_1},\mu_{i_2},\cdots ,\mu_{i_N} ) = \prod_{\substack{j=1\\m_j \leq N-2 }}^{N/2 -1} (-1)^{m_j} \,.
\end{equation}
This explicitly gives
\begin{equation}
\boxed{
\langle 0| \mathrm{T}\gamma^{\mu_1}\gamma^{\mu_2}\cdots \gamma^{\mu_N}|0\rangle =\frac{1}{D^{N/2}} \sum_{\substack{i_1 ,i_2 ,\cdots, i_N \in S \\i_1 < i_2 <\cdots < i_N }} \prod_{\substack{j=1\\m_j \leq N-2 }}^{N/2 -1} (-1)^{m_j} \langle 0|\gamma^{\mu_{i_a}}\cdots \gamma^{\mu_{i_l}} |0\rangle}\,.
\end{equation}
This evaluates to give
\begin{equation}
\begin{aligned}
\langle 0| \mathrm{T}\gamma^{\mu_1}\gamma^{\mu_2}\cdots \gamma^{\mu_N}|0\rangle &=\frac{1}{D^{N/2}} \sum_{\substack{i_1 ,i_2 ,\cdots, i_N \in S \\i_1 < i_2 <\cdots < i_N }} \prod_{\substack{j=1\\m_j \leq N-2 }}^{N/2 -1} (-1)^{m_j} \prod_{\substack{i_a , i_b =1,2,\cdots , N \\i_a < i_b}}^{N/2}\langle 0|\gamma^{\mu_{i_a}}\gamma^{\mu_{i_b}}|0\rangle \\
&= \sum_{\substack{i_1 ,i_2 ,\cdots, i_N \in S \\i_1 < i_2 <\cdots < i_N }} \prod_{\substack{j=1\\m_j \leq N-2 }}^{N/2 -1} (-1)^{m_j} \prod_{\substack{i_a , i_b =1,2,\cdots , N \\ i_a < i_b}}^{N/2} \eta^{\mu_{i_a}\mu_{i_b}}\,.
\end{aligned}
\end{equation}
Let's illustrate this by an example, let's say $N= 6$ with the 6-point correlation function of $\langle 0| \mathrm{T}\gamma^{\mu_1}\gamma^{\mu_2} \gamma^{\mu_3}\gamma^{\mu_4}\gamma^{\mu_5}\gamma^{\mu_6}|0\rangle $.  There are a total of 15 possible contractions. For example, by mathematical computation we have a term of
\begin{equation}
-\eta^{\mu_1\mu_4}\eta^{\mu_2\mu_5}\eta^{\mu_3\mu_6}\,.
\end{equation}
Now we want to recover the minus sign.
We have
\begin{equation}
\begin{aligned}
&\quad\wick{
\frac{1}{D^3}\langle 0| \c1{\gamma}^{\mu_{i_1}} \c3{\gamma}^{\mu_{i_2}} \c4{\gamma}^{\mu_{i_3}} \c1{\gamma}^{\mu_{i_4}}\c3{\gamma}^{\mu_{i_5}} \c4{\gamma}^{\mu_{i_6}} |0\rangle 
}\\
&= \frac{1}{D^3}(-1)^2\langle 0|{\gamma}^{\mu_{i_1}}{\gamma}^{\mu_{i_4}}{\gamma}^{\mu_{i_2}}{\gamma}^{\mu_{i_3}}{\gamma}^{\mu_{i_5}}{\gamma}^{\mu_{i_6}}|0\rangle\\
&= \frac{1}{D^3}(-1)^2 (-1)^1 \langle 0|{\gamma}^{\mu_{i_1}}{\gamma}^{\mu_{i_4}} {\gamma}^{\mu_{i_2}}{\gamma}^{\mu_{i_5}}{\gamma}^{\mu_{i_3}}{\gamma}^{\mu_{i_6}}|0\rangle\\
&= -\eta^{\mu_1\mu_4}\eta^{\mu_2\mu_5}\eta^{\mu_3\mu_6}\,.
\end{aligned}
\end{equation}

\section{Geometrical Formalism}
In this section, we will give a mathematical formalism for the gamma quantum fields in geometrical means. 
First, we know that by the definition of p-form, it reads
\begin{equation}
\alpha = \frac{1}{p!}a_{\mu_1 \mu_2 \cdots \mu_p}\omega^{\mu_1}\wedge \omega^{\mu_2}\wedge \cdots \wedge \omega^{\mu_p}
\end{equation}
where $a_{\mu_1 \mu_2 \cdots \mu_p}$ is an totally anti-symmetric tensor. 

In analogy, we construct a quantized p-form formed by the gamma quantum fields. We will drop the hat notation as for convenience,
\begin{equation}
\alpha = \frac{1}{p!}a_{\mu_1 \mu_2 \cdots \mu_p} \gamma^{\mu_1}\gamma^{\mu_2}\cdots \gamma^{\mu_n}
\end{equation}
In particular, for $a_{\mu_1 \mu_2 \cdots \mu_p} =\epsilon_{\mu_1 \mu_2 \cdots \mu_p}$ the Levi-civita tensor for $p=4$ it follows that in fact the $\gamma^5$ matrix is a matrix 4-form as follows
\begin{equation}
-i\gamma^5 = \frac{1}{4!}\epsilon_{\mu_1 \mu_2 \mu_3 \mu_4}\gamma^{\mu_1}\gamma^{\mu_2}\gamma^{\mu_3} \gamma^{\mu_n}
\end{equation}
Next we have the following theorem for forms. Suppose $\alpha$ is a $q$-form and $\beta$ is a $p$-form, then
\begin{equation}
\alpha\wedge \beta = (-1)^{pq}\beta\wedge\alpha \,.
\end{equation}
For the case of gamma matrices, we have
\begin{equation}
\alpha\beta = (-1)^{pq}\beta\alpha\,.
\end{equation}
In explicit tensor notation,
\begin{equation}
\alpha_{ab}\beta_{bc} = (-1)^{pq}\beta_{ab}\alpha_{bc}\,.
\end{equation}

Now consider $\alpha$ as a $p$-form and $\beta$ as a $q$-form. Let 
\begin{equation}
\alpha = \frac{1}{p!}a_{\mu_1 \mu_2 \cdots \mu_p}\omega^{\mu_1}\wedge \omega^{\mu_2}\wedge \cdots \wedge \omega^{\mu_p} \quad\text{and}\quad
\beta=\frac{1}{q!}b_{\mu_{p+1}\mu_{p+2}\cdots \mu_{p+q} }\omega^{\mu_{p+1}}\wedge \omega^{\mu_{p+2}}\wedge\cdots \wedge \omega^{\mu_{p+q}}\,. 
\end{equation}
Then
\begin{equation}
\begin{aligned}
\alpha\wedge\beta &=\frac{1}{p!q!}a_{\mu_1 \mu_2 \cdots \mu_p}b_{\mu_{p+1}\mu_{p+2}\cdots \mu_{p+q} }\omega^{\mu_1}\wedge \omega^{\mu_2}\wedge \cdots \wedge \omega^{\mu_p} \wedge \omega^{\mu_{p+1}}\wedge \omega^{\mu_{p+2}}\wedge\cdots \wedge \omega^{\mu_{p+q}} \\
&=\frac{(p+q)!}{p!q!} a_{[\mu_1 \mu_2 \cdots \mu_p}b_{\mu_{p+1}\mu_{p+2}\cdots \mu_{p+q}]}\omega^{\mu_1}\wedge \omega^{\mu_2}\wedge \cdots \wedge \omega^{\mu_p} \wedge \omega^{\mu_{p+1}}\wedge \omega^{\mu_{p+2}}\wedge\cdots \wedge \omega^{\mu_{p+q}} \\
&\equiv (\alpha\wedge\beta)_{\mu_1 \mu_2 \cdots \mu_p\mu_{p+1}\mu_{p+2}\cdots \mu_{p+q} }\omega^{\mu_1}\wedge \omega^{\mu_2}\wedge \cdots \wedge \omega^{\mu_p} \wedge \omega^{\mu_{p+1}}\wedge \omega^{\mu_{p+2}}\wedge\cdots \wedge \omega^{\mu_{p+q}}
\end{aligned}
\end{equation}
Therefore we have 
\begin{equation}
(\alpha\wedge\beta)_{\mu_1 \mu_2 \cdots \mu_p\mu_{p+1}\mu_{p+2}\cdots \mu_{p+q} } = \frac{(p+q)!}{p!q!} a_{[\mu_1 \mu_2 \cdots \mu_p}b_{\mu_{p+1}\mu_{p+2}\cdots \mu_{p+q}]}
\end{equation}
In terms of gamma field basis
\begin{equation}
\alpha\beta= (\alpha\beta)_{\mu_1 \mu_2 \cdots \mu_p\mu_{p+1}\mu_{p+2}\cdots \mu_{p+q} }\gamma^{\mu_1} \gamma^{\mu_2} \cdots  \gamma^{\mu_p}  \gamma^{\mu_{p+1}} \gamma^{\mu_{p+2}}\cdots  \gamma^{\mu_{p+q}} 
\end{equation}

Now we will study a very important concept on hodge dual. Suppose $\alpha$ is a $p$-form, The definition of its dual is given by \cite{Hervik}
\begin{equation}
\star\alpha =\frac{1}{p!(n-p)!}\epsilon_{\nu_1\cdots\nu_p\mu_1\cdots\mu_{n-p}}a^{\nu_1\cdots \nu_p}\omega^{\mu_1}\wedge\cdots\wedge \omega^{\mu_{n-p}} \,.
\end{equation}
such that $\star\alpha$ is a $n-p$ form. Next, by definition,
\begin{equation}
\star(\omega^{\nu_1}\wedge\cdots\wedge\omega^{\nu_p}) = \frac{1}{(n-p)!}\sqrt{|g|} |g_p|^{-1}\epsilon_{\nu_1\cdots\nu_p\mu_1\cdots\mu_{n-p}}\omega^{\mu_1}\wedge \cdots \wedge \omega^{\mu_{n-p}}
\end{equation} 
where $g_p$ is the determinant of the metric tensor associated with the space of the $p$-form of $\alpha$, and $g$ is the determinant of the metric in the $n$-dimensional space. 
For our case, we have
\begin{equation}
\star\alpha =\frac{1}{p!(n-p)!}\epsilon_{\nu_1\cdots\nu_p\mu_1\cdots\mu_{n-p}}a^{\nu_1\cdots \nu_p}\gamma^{\mu_1}\cdots \gamma^{\mu_{n-p}} \,.
\end{equation}
and
\begin{equation}
\star(\gamma^{\nu_1}\cdots\gamma^{\nu_p}) = \frac{1}{(n-p)!}\sqrt{|g|} |g_p|^{-1}\epsilon_{\nu_1\cdots\nu_p\mu_1\cdots\mu_{n-p}}\gamma^{\mu_1} \cdots  \gamma^{\mu_{n-p}}
\end{equation} 
The metric is defined by
\begin{equation}
ds^2 = g_{\mu\nu}\gamma^{\mu}\gamma^{\nu} 
\end{equation}
for which
\begin{equation}
g_{\mu\nu} =\frac{1}{4}\mathrm{Tr}(\gamma_{\mu}\gamma_{\nu})=\frac{1}{4}\sum_{ab}\gamma_{\mu ab}\gamma_{\nu ba}\,.
\end{equation}
It is easy to see that the factor $\sqrt{|g|} |g_p|^{-1} =1$. Now consider $p=1$ case, then
\begin{equation}
\star\gamma^{\nu_1} =\frac{1}{(4-1)!}(1) \epsilon_{\nu_1\mu_1\mu_2\mu_3}\gamma^{\mu_1}\gamma^{\mu_2}\gamma^{\mu_3}\,.
\end{equation}
We then have
\begin{equation}
\star \gamma^0 = \frac{1}{3!} (3!)\epsilon_{0123}\gamma^1\gamma^2\gamma^3 \,.
\end{equation}
Therefore we have
\begin{equation}
\boxed{
\star\gamma^0 = \gamma^1\gamma^2\gamma^3 =-i\gamma^0\gamma^5 \,.
}
\end{equation}
Next,
\begin{equation}
\star\gamma^1 = \epsilon_{1023} \gamma^0\gamma^2\gamma^3
\end{equation}
Thus
\begin{equation}
\boxed{
\star\gamma^1 =-\gamma^0\gamma^2\gamma^3=-i\gamma^1\gamma^5 \,.
}
\end{equation}
Then
\begin{equation}
\star\gamma^2 = \epsilon_{2013}\gamma^0\gamma^1\gamma^3 \,.
\end{equation}
Thus
\begin{equation}
\boxed{
\star\gamma^2 = \gamma^0\gamma^1 \gamma^3 = -i\gamma^2\gamma^5}
\end{equation}
Finally
\begin{equation}
\star\gamma^3 =\epsilon_{3012}\gamma^0\gamma^1\gamma^2 \,.
\end{equation}
Therefore,
\begin{equation}
\boxed{
\star\gamma^3 =-\gamma^0\gamma^1\gamma^2 =-i\gamma^3\gamma^5 \,.
}
\end{equation}
In other words we have
\begin{equation}
\boxed{
\star\gamma^\mu = -i\gamma^\mu\gamma^5
}
\end{equation}
Since $\gamma^\mu$ is a vector field and $\gamma^\mu\gamma^5$ is an axial vector field, we see that the dual of the vector field is the axial vector field. And by $\{\gamma^5,\gamma^\mu\}=0$, we have
\begin{equation}
\star\gamma^\mu = i\gamma^5\gamma^\mu
\end{equation}
Therefore we have the hodge dual operator as
\begin{equation}
\boxed{
\star = i\gamma^5
}
\end{equation}
As from $\star\gamma^0 = \gamma^1\gamma^2\gamma^3$, it follows that
\begin{equation}
\gamma^0 \star \gamma^0 = \gamma^0\gamma^1\gamma^2\gamma^3
\end{equation}
It follows that
\begin{equation}
\gamma^5 =i\gamma^0 \star\gamma^0
\end{equation}
In general, we find
\begin{equation}
\boxed{\gamma^5 = i\gamma^\mu \star \gamma^\mu} 
\end{equation}
Since \begin{equation}
\gamma^0\gamma^1 \gamma^2\gamma^3 = \gamma^3\gamma^2\gamma^0\gamma^1,
\end{equation}
therefore we find that $\gamma^5$ is a dual invariant in 4 dimension.
In general
\begin{equation}
\gamma^0\gamma^1 \cdots \gamma^n = (-1)^{\frac{n(n+1)}{2}} \gamma^n \cdots\gamma^1\gamma^0\,.
\end{equation}
Next, we can promote to other basis. Notice that
\begin{equation}
\star(\gamma^{\nu_1}\gamma^{\nu_2}) = \frac{1}{(4-2)!}\sqrt{|g|}|g_p |^{-1} \epsilon_{\nu_1\nu_2\mu_1\mu_2 }\gamma^{\mu_1} \gamma^{\mu_2} \,.
\end{equation}
Then for example
\begin{equation}
\begin{aligned}
\star(\gamma^0\gamma^1) &=\frac{1}{2!}(1) \epsilon_{01\mu_1\mu_2}\gamma^{\mu_1} \gamma^{\mu_2}\\
&=\frac{1}{2!}(2!)(1) \epsilon_{0123}\gamma^{2}\gamma^3 \,.
\end{aligned}
\end{equation}
Thus
\begin{equation}
\star(\gamma^{0}\gamma^1)=\gamma^2\gamma^3\,.
\end{equation}
Since the charge conjugation is given by the operator $C=i\gamma^2\gamma^0$ and the time reversal operator is given by $T=i\gamma^1\gamma^3$. We find that 
\begin{equation}
\star(\gamma^2\gamma^0) =\epsilon_{2013}\gamma^1\gamma^3 = \gamma^1\gamma^3 \,.
\end{equation}
Therefore we prove that the charge conjugation and the time reversal operator is dual to each other
\begin{equation}
\boxed{
\star C = T 
}
\end{equation}
Finally, we would like to find out the dual of the $\gamma^5$ operator. By definition,
\begin{equation}
\star \gamma^5 =\star i \star\gamma^0 \star\gamma^1 \star \gamma^2 \star\gamma^3\,.
\end{equation}
Explicitly, using the above results , we have
\begin{equation}
\begin{aligned}
\star\gamma^5 &= (\pm 1)(\gamma^1\gamma^2\gamma^3)(-\gamma^0\gamma^2\gamma^3)(\gamma^0\gamma^1\gamma^3)(-\gamma^0\gamma^1 \gamma^2) \\
&=(\pm 1)(-1)^3 (-1)^2 (-1)^3 \gamma^0\gamma^1\gamma^2\gamma^3\,\gamma^3\gamma^2\gamma^1 \gamma^0 \,\gamma^0\gamma^1\gamma^2\gamma^3\\
&=(\pm 1)\gamma^0\gamma^1\gamma^2\gamma^3\,\gamma^0\gamma^1\gamma^2\gamma^3\,\gamma^0\gamma^1\gamma^2\gamma^3\\
&=\pm 1 (-i\gamma^5)^3 \\
&= \pm i\gamma^5
\end{aligned}
\end{equation}
Therefore, we have the following eigen-matrix equation
\begin{equation}
\star\gamma^5 = \pm i \gamma^5
\end{equation}
Therefore, we find that in fact $\gamma^5$ is the eigen-matrix of the hodge dual operator.
Next we find
\begin{equation}
\star\gamma^5 \star \gamma^5 =  \star(\gamma^5)^\dagger \star \gamma^5 =\pm i \pmb{1}\,.
\end{equation}

With the definition of dual vector field as axial vector field, now we are interested in calculating the axial vector field strength. First we would like to compute the Dirac algebra of axial vector field:
\begin{equation}
\begin{aligned}
\{ \star\gamma^\mu ,\star\gamma^\nu\}&= \{-i\gamma^\mu \gamma^5 ,-i\gamma^\mu\gamma^5\}\\
&= -(\gamma^\mu\gamma^5\gamma^\nu\gamma^5 + \gamma^\nu\gamma^5\gamma^\mu\gamma^5 ) \\
&= (\gamma^\mu \gamma^5\gamma^5\gamma^\nu +\gamma^\nu \gamma^5\gamma^5\gamma^\mu  )\\
&=\{\gamma^\mu\gamma^\nu +\gamma^\nu\gamma^\mu \} \\
&=\{\gamma^\mu, \gamma^\nu\}\\
&=2\eta^{\mu\nu}I\,.
\end{aligned}
\end{equation}
Hence the axial field and vector field both share the same Dirac algebra.
Now consider the dual axial gauge field strength $\star F_{\mu\nu}$:
\begin{equation}
\star F = d\star\gamma + \star\gamma\wedge\star\gamma \,.
\end{equation}
Explicitly,
\begin{equation}
\begin{aligned}
\star F^{\mu\nu} &= \partial^{\mu}\star\gamma^{\nu} -\partial^{\nu}\star\gamma^{\mu} +\star\gamma^{\mu}\star\gamma_{\nu} - \star\gamma^{\nu}\star\gamma^{\mu} \\
&= -(\gamma^\mu\gamma^5\gamma^\nu\gamma^5 - \gamma^\nu\gamma^5\gamma^\mu\gamma^5) \\
&=\gamma^\mu\gamma^\nu - \gamma^\nu\gamma^\mu\\
&= -4iS^{\mu\nu}
\end{aligned}
\end{equation}
Therefore we have
\begin{equation}
\star F^{\mu\nu} = F^{\mu\nu}
\end{equation}
and simply
\begin{equation}
\boxed{
\star F = F
}\,.
\end{equation}
Hence $F$ is a dual invariant. It follows that the Lagrangian density is 
\begin{equation}
\mathcal{L}=\star F_{\mu\nu}\star F^{\mu\nu} = F_{\mu\nu}F^{\mu\nu}\,.
\end{equation}

\subsection{Dirac gamma representation of EM-field tensor}
Notice that the object $\gamma^{\mu}\gamma^{\nu}$ can be decomposed into symmetric part and anti-symmetric part, where the former is the symmetric metric tensor and the latter is proportional to the anti-symmetric tensor,
\begin{equation}
\gamma^\mu \gamma^\nu = \frac{1}{2}(\gamma^{\mu}\gamma^{\nu} + \gamma^{\nu}\gamma^{\mu}) + \frac{1}{2}(\gamma^{\mu}\gamma^{\nu} - \gamma^{\nu}\gamma^{\mu}) = \eta^{\mu\nu} -i\sigma^{\mu\nu} \,.
\end{equation}
Next we construct the field strength tensor by Dirac matrices as follow:
\begin{equation}
F^{\mu\nu} = \frac{1}{2}(\gamma^{\mu}\gamma^{\nu} - \gamma^{\nu}\gamma^{\mu}) =\begin{pmatrix}
0 &\gamma^0\gamma^1 &\gamma^0\gamma^2 & \gamma^0 \gamma^3 \\
-\gamma^0\gamma^1 & 0 & \gamma^1\gamma^2 & \gamma^1\gamma^3\\
-\gamma^0\gamma^2  & -\gamma^1\gamma^2 &0& \gamma^2\gamma^3 \\-\gamma^0\gamma^3 &-\gamma^1\gamma^3 &-\gamma^2\gamma^3 &0\\
\end{pmatrix}
\end{equation}
But as we find that \begin{equation}
\gamma^1\gamma^2 =\star (\gamma^0\gamma^3) , \quad \gamma^1\gamma^3 =-\star (\gamma^0\gamma^2)  \quad\text{and}\quad \gamma^2\gamma^3 = \star(\gamma^0\gamma^1) 
\end{equation}
Therefore we find that the field strength tensor by Dirac matrices as
\begin{equation}
F^{\mu\nu} = \begin{pmatrix}
0 &\gamma^0\gamma^1 &\gamma^0\gamma^2 & \gamma^0 \gamma^3 \\
-\gamma^0\gamma^1 & 0 & \star(\gamma^0\gamma^3) & -\star(\gamma^0\gamma^2)\\
-\gamma^0\gamma^2  & -\star(\gamma^0\gamma^3) &0& \star(\gamma^0\gamma^1) \\-\gamma^0\gamma^3 & \star(\gamma^0\gamma^2) &-\star(\gamma^0\gamma^1) &0\\
\end{pmatrix}
\end{equation}
This has exactly the same form with the EM-field tensor, which is, under the metric equal to $(-+++)$convention
\begin{equation}
F^{\mu\nu}=
\begin{pmatrix}
0 & E_x &E_y &E_z \\
-E_x & 0 & B_z & -B_y\\
-E_y & -B_z & 0 & B_x \\
-E_z & B_y & -B_x & 0
\end{pmatrix}
\end{equation}
where we identify
\begin{equation}
E_x = \gamma^0\gamma^1 , \quad E_y = \gamma^0\gamma^2 ,\quad E_z = \gamma^0\gamma^3
\end{equation}
and
\begin{equation}
B_x =\star(\gamma^0 \gamma^1),\quad B_y=-\star(\gamma^0 \gamma^2),\quad B_z =\star(\gamma^0 \gamma^3) \,.
\end{equation}
It directly follows that
\begin{equation}
\boxed{
B_x =\star E_x ,\quad B_y = -\star E_y \quad\text{and}\quad B_z =\star E_z }\,.
\end{equation}
Therefore we proved that electric field and magnetic field are hodge dual to each other, this result makes sense as magnetic field is an axial vector while electric field is the polar vector, and axial vector and polar vector are dual to each other.

Define the normal direct sum $\oplus $ as right direct sum  $\oplus_R$ as the sum along the diagonal blocks, and the define the left direct sum $\oplus_L$ as the sum along the off-diagonal blocks. Under the Dirac representation, explicitly we have
\begin{equation}
E_x=\gamma^0 \gamma^1 =\begin{pmatrix}
 0 & \sigma_x \\
 \sigma_x & 0 \\
\end{pmatrix}, \quad  E_y=\gamma^0 \gamma^2 =\begin{pmatrix}
 0 & \sigma_y \\
 \sigma_y & 0 \\
\end{pmatrix}, \quad E_z =  \gamma^0 \gamma^3 =\begin{pmatrix}
 0 & \sigma_z \\
 \sigma_z & 0 \\
\end{pmatrix} \,.
\end{equation} 
Therefore, for electric fields, we have
\begin{equation}
E^i = \gamma^0\gamma^i = \sigma^i \oplus_L \sigma^i
\end{equation}
And for magnetic field,
\begin{equation}
B_x=\gamma^2 \gamma^3 =\begin{pmatrix}
 -i\sigma_x & 0   \\
 0 & -i\sigma_x \\
\end{pmatrix}, \quad  B_y=-\gamma^1 \gamma^3 =\begin{pmatrix}
i \sigma_y & 0\\
 0  & i\sigma_y \\
\end{pmatrix}, \quad B_z = \gamma^1 \gamma^2 =\begin{pmatrix}
-i\sigma_z & 0 \\
0 & -i\sigma_z \\
\end{pmatrix} \,.
\end{equation} 
Thus we have
\begin{equation}
B^j = (-1)^{j+1}i(\sigma^j \oplus_R \sigma^j)\,.
\end{equation}
Then it follows that
\begin{equation}
\begin{cases}
-i(\sigma_x \oplus_R \sigma_x) &= \star(\sigma_x \oplus_L \sigma_x)\\
-i(\sigma_y \oplus_R \sigma_y) &= -\star(\sigma_y \oplus_L \sigma_y)\\
-i(\sigma_z \oplus_R \sigma_x) &= \star(\sigma_z \oplus_L \sigma_z)
\end{cases}
\end{equation}
Therefore we see that
\begin{equation}
\boxed{
\star \oplus_L =\pm i \oplus_R 
} \,.
\end{equation}
The basically infers that left direct sum is hodge dual to right direct sum up to a factor of $\pm i$.

\section{Integral Formalism for Gamma fields}
Since now we know that gamma field behaves as both a gauge field and a fermionic field, it is straight forward to study its integral formalism. For non-abelian gauge fields, according to \cite{Yang2}, a  path dependent gauge field phase  factor from $A=X$ to $B=X+dx$ is given by
\begin{equation}
\phi_{AB}= \phi_{X(X+dx)} = I + A^a_{\mu}(t^a)^{i}_{\,\,\,j} dx^\mu \equiv A_{\mu\,\,\,j}^{\,\,i}dx^\mu
\end{equation} 
Furthermore, it is observed that the non-Abelian gauge field is traceless,
\begin{equation} \label{eq:157}
\mathrm{Tr}\,A_{\mu} = 0\,.
\end{equation}
And now consider a small loop ABCDA, then the phase factor becomes
\begin{equation}
\phi_{ABCDA} = I + F_{\mu\nu}^a (t^a)^{i}_{\,\,\,j}dx^\mu \wedge dx^\nu \equiv I + F_{\mu\nu\,\,\,j}^{\,\,\,\,\,i}dx^\mu \wedge dx^\nu \,,
\end{equation}
where
\begin{equation}
F_{\mu\nu\,\,\,j}^{\,\,\,\,\,i} =\partial_{\mu}A_{\nu\,\,\,j}^{\,\,i} -\partial_{\nu}A_{\mu\,\,\,j}^{\,\,i} +  A_{\mu\,\,\,k}^{\,\,i}A_{\nu\,\,\,j}^{\,\,k} - A_{\nu\,\,\,k}^{\,\,i}A_{\mu\,\,\,j}^{\,\,k}
\end{equation}
is the curvature tensor. According differential geometry, the curvature is defined by
\begin{equation}
\Omega^{i}_{\,\,\,j} = \frac{1}{2} F_{\mu\nu\,\,\,j}^{\,\,\,\,\,i}dx^\mu \wedge dx^\nu \,,
\end{equation}
and as we know that the first Chern class is defined by
\begin{equation}
c_1 = \frac{i}{2\pi} \mathrm{Tr}\, \Omega \,,
\end{equation}
Since the SU(N) Lie group generator $t^a$ is traceless, then $F_{\mu\nu\,\,\,i}^{\,\,\,\,\,i} =0$, therefore, the SU(N) lie group is a trivial principal bundle on the manifold. 

Now we promote this idea to study the case of gamma fields. The phase is given by
\begin{equation}
\phi_{AB}= \phi_{X(X+dx)} = I + \gamma_{\mu\,\,\,b}^{\,\,a} dx^\mu
\end{equation}
and notice that similar to \ref{eq:157}, this amounts to give 
\begin{equation}
\mathrm{Tr}\,\gamma_{\mu} = 0 \,.
\end{equation}
Then for a differential loop, we have 
\begin{equation}
\phi_{ABCDA} = I + s_{\mu\nu\,\,\,b}^{\,\,\,\,\,a} dx^\mu dx^\nu \,,
\end{equation}
where
\begin{equation}
 s_{\mu\nu\,\,\,b}^{\,\,\,\,\,a} =\partial_{\mu}\gamma_{\nu\,\,\,b}^{\,\,a} -\partial_{\nu}\gamma_{\mu\,\,\,b}^{\,\,a} +  \gamma_{\mu\,\,\,c}^{\,\,a}\gamma_{\nu\,\,\,b}^{\,\,c} - \gamma_{\nu\,\,\,c}^{\,\,a}\gamma_{\mu\,\,\,b}^{\,\,c} = \gamma_{\mu\,\,\,c}^{\,\,a}\gamma_{\nu\,\,\,b}^{\,\,c} - \gamma_{\nu\,\,\,c}^{\,\,a}\gamma_{\mu\,\,\,b}^{\,\,c} \,.
\end{equation}
The curvature is defined by 
\begin{equation}
\Omega^{i}_{\,\,\,j} = \frac{1}{2} s_{\mu\nu\,\,\,j}^{\,\,\,\,\,i}dx^\mu \wedge dx^\nu \,,
\end{equation}
But since $s_{\mu\nu\,\,\,a}^{\,\,\,\,\,a}=\mathrm{Tr}\,(\gamma^\mu \gamma^\nu - \gamma^\nu \gamma^\mu ) =0$, so this also yields vanishing first chern class. 
  To relate the above result to the case of general relativity, we know that the curvature is defined by
\begin{equation}
R_{\mu\nu\,\,\,\sigma}^{\,\,\,\,\,\rho} =\partial_{\mu}\Gamma_{\nu\,\,\,\sigma}^{\,\,\rho} -\partial_{\nu}\Gamma_{\mu\,\,\,\sigma}^{\,\,\rho} +  \Gamma_{\mu\,\,\,\lambda}^{\,\,\rho}\Gamma_{\nu\,\,\,\sigma}^{\,\,\lambda} - \Gamma_{\nu\,\,\,\lambda}^{\,\,\rho}\Gamma_{\mu\,\,\,\sigma}^{\,\,\lambda}
\end{equation}
The curvature is defined by
\begin{equation}
\Omega^{\rho}_{\,\,\,\sigma} = \frac{1}{2} R_{\mu\nu\,\,\,\sigma}^{\,\,\,\,\,\rho}dx^\mu \wedge dx^\nu \,.
\end{equation}
Notice that the trace of the Christoffel connection yields,
\begin{equation}
\mathrm{Tr}\,\Gamma_{\mu} =\Gamma_{\mu\,\,\,\rho}^{\,\,\rho} = 
\partial_{\mu}\ln\sqrt{-g}\,.
\end{equation}
Therefore, the first chern class is given by
\begin{equation}
c_1 = \frac{1}{2\pi} ( \Gamma_{\mu\,\,\,\lambda}^{\,\,\rho}\Gamma_{\nu\,\,\,\rho}^{\,\,\lambda} - \Gamma_{\nu\,\,\,\lambda}^{\,\,\rho}\Gamma_{\mu\,\,\,\rho}^{\,\,\lambda})dx^\mu \wedge dx^\nu \,.
\end{equation}
Furthermore, the chern number, which is defined by the integral of curvature 2-form, defines the topological invariant winding number 
\begin{equation}
c = \frac{1}{2\pi}\int ( \Gamma_{\mu\,\,\,\lambda}^{\,\,\rho}\Gamma_{\nu\,\,\,\rho}^{\,\,\lambda} - \Gamma_{\nu\,\,\,\lambda}^{\,\,\rho}\Gamma_{\mu\,\,\,\rho}^{\,\,\lambda})dx^\mu \wedge dx^\nu \,.
\end{equation}
We would like to study the case when the trace of the connection is zero,  i,e, $\mathrm{Tr}\,\Gamma_{\mu} = 
\partial_{\mu}\ln\sqrt{-g}=0$. Thus $\ln\sqrt{-g} =c$. Then $g_{\mu\nu} =c_{\mu\nu}$ is a constant metric. Then, traceless Christoffel connection amounts to vanishing curvature, thus vanishing Chern class and Chern number. Thus the trivial gauge bundle for $A_{\mu}$ and $\gamma_{\mu}$ corresponds to the flat metric.

\section{Conclusion}
In this short article, we provide a completely new perspective on $\gamma^\mu$ matrix, for which can be considered as both a global gauge field and fermonic field, quantization of gamma field will be new particles, which is the nature intrinsic fifth force. Wick's theorem of gamma fields is established and is just the normal trace algebra of gamma matrices. Next, we have constructed a formalism for gamma matrices in the notion of differential geometry. Then we find that the electromagnetic field tensor can be isomorphically expressed in-terms of the gamma matrix representations. Finally, we study the integral formalism of Gamma fields.

\end{document}